\documentclass[10pt,conference]{IEEEtran}
\IEEEoverridecommandlockouts

\usepackage{cite}
\usepackage{amsmath,amssymb,amsfonts}
\usepackage{algorithmic}
\usepackage{graphicx}
\usepackage{textcomp}
\usepackage{xcolor}
\usepackage{multirow}
\usepackage{comment}
\usepackage{balance}
\usepackage{hyperref}

\usepackage{svg}

\usepackage{xcolor}
\usepackage{todonotes}

\def\BibTeX{{\rm B\kern-.05em{\sc i\kern-.025em b}\kern-.08em
    T\kern-.1667em\lower.7ex\hbox{E}\kern-.125emX}}
\begin{document}

\title{Music Genre Classification using Large Language Models
\thanks{Identify applicable funding agency here. If none, delete this.}
}

\author{\IEEEauthorblockN{Mohamed El Amine Meguenani}
\IEEEauthorblockA{\textit{École de Technologie Supérieure} \\
\textit{University of Quebec}\\
Montréal, Canada \\
mohamed-el-amine.meguenani.1@ens.etsmtl.ca}
\and
\IEEEauthorblockN{Alceu de Souza Britto Jr.}
\IEEEauthorblockA{\textit{Prog. de Pósgraduação em Informática} \\
\textit{Pontifícia Univ. Católica do Paraná}\\
Curitiba, Brazil \\
alceu.britto@pucpr.br}
\and
\IEEEauthorblockN{Alessandro Lameiras Koerich}
\IEEEauthorblockA{\textit{École de Technologie Supérieure} \\
\textit{University of Quebec}\\
Montréal, Canada \\
alessandro.koerich@etsmtl.ca}
}

\maketitle

\begin{abstract}
This paper exploits the zero-shot capabilities of pre-trained large language models (LLMs) for music genre classification. The proposed approach splits audio signals into 20ms chunks and processes them through convolutional feature encoders, a transformer encoder, and additional layers for coding audio units and generating feature vectors. The extracted feature vectors are used to train a classification head. During inference, predictions on individual chunks are aggregated for a final genre classification. We conducted a comprehensive comparison of LLMs, including WavLM, HuBERT, and wav2vec 2.0, with traditional deep learning architectures like 1D and 2D convolutional neural networks (CNNs) and the audio spectrogram transformer (AST). Our findings demonstrate the superior performance of the AST model, achieving an overall accuracy of 85.5\%, surpassing all other models evaluated. These results highlight the potential of LLMs and transformer-based architectures for advancing music information retrieval tasks, even in zero-shot scenarios.
\end{abstract}


\section{Introduction}
Music is an integral component of human culture, universally present across diverse societies and civilizations \cite{Lidy2010}. Its significance extends far beyond mere entertainment, serving as a powerful tool for enhancing mental and emotional well-being \cite{REBECCHINI2021}. Research has demonstrated that individuals engage with music for various purposes, including improving concentration, mitigating stress, and alleviating negative emotional states such as loneliness and sadness \cite{Prabhakar2023}. This multifaceted role of music in the human experience underscores its importance not only as a cultural artifact but also as a potential therapeutic intervention.

The exponential growth of digital music libraries in recent decades has necessitated sophisticated methods for organizing, searching, and understanding musical data, leading to the emergence of the music information retrieval (MIR) field \cite{Tzanetakis2002}. As a subfield of music technology, MIR focuses on developing algorithms, systems, and applications for extracting, analyzing, processing, and retrieving information from various musical data sources, including audio signals, symbolic representations, and metadata \cite{Schedl2014}. This interdisciplinary field has experienced remarkable growth over the past two decades, driven by technological advancements in audio compression techniques and increased computing power. These developments have facilitated more efficient manipulation of musical data, opening up new avenues for research and application \cite{Burgoyne2015}.

MIR encompasses a wide range of tasks, from feature extraction to creating personalized, contextual, and adaptive systems. Among these, music genre classification (MGC) has emerged as a particularly significant area of focus \cite{Tzanetakis2002}. Genre classification plays a crucial role in MIR by enhancing the organization and efficiency of information retrieval, thereby facilitating music discovery and recommendation. However, this task presents unique challenges due to its reliance on subtle characteristics, human and cultural subjectivity, and the need to discern cultural and social aspects from sound elements alone \cite{Lidy2010}. Furthermore, determining the specific characteristics for effective music classification remains a complex issue \cite{Allamy2021}.

The development of an effective MIR system capable of accurately classifying music genres has the potential to revolutionize the music industry by improving accessibility and enhancing user experience \cite{Prabhakar2023}. As such, addressing the challenges associated with genre classification represents a critical frontier in MIR research, with implications for both academic understanding and practical applications in the music technology landscape.

To address the challenges and enhance the accuracy of MGC, a critical task in MIR, this study explores the application of large language models (LLMs). These models have demonstrated remarkable success in various speech-related tasks \cite{Yang2021}, suggesting potential for adaptation to music analysis. Specifically, audio LLMs such as HuBERT \cite{Hsu2021}, wav2vec 2.0 \cite{Baevski2020}, and WaveLM \cite{Chen2022} have achieved state-of-the-art performance in diverse audio processing domains, including phoneme classification, speaker identification and verification, emotion recognition, speech translation, and spoken language understanding \cite{Yang2021}. The proven efficacy of these models in capturing complex audio features and patterns presents a promising avenue for improving MGC accuracy.

This paper aims to evaluate the efficacy of recent audio LLMs, specifically HuBERT, wav2vec 2.0, and WaveLM, in the context of MGC. These models have demonstrated state-of-the-art performance across various audio tasks \cite{Yang2021}, prompting our investigation into their potential for music analysis. We hypothesize that these LLMs, even in a zero-shot scenario, can identify complex patterns in musical signals and offer multifaceted approaches to the challenging task of MGC. Our research methodology encompasses a comprehensive analysis of the internal behavior of these models, with a focus on identifying the layers that provide optimal representations for MGC \cite{Pasad2023}. Additionally, we conduct a comparative analysis, bringing together the performance of LLM-based approaches and traditional convolutional and transformer models that do not rely on language modeling techniques. The novelty of this work lies in the application of LLMs to MGC, a previously unexplored domain for these models. By establishing a baseline for LLM performance in this context, we aim to pave the way for future research that leverages LLMs for a broader range of music-related tasks. This study not only contributes to the advancement of MIR techniques but also expands the potential applications of LLMs beyond their original speech-related domains.

The remainder of this paper is structured as follows: Section \ref{sec:rel} provides a comprehensive review of related works, contextualizing our research within the existing literature on MGC and the application of LLMs to audio tasks. Section \ref{sec:appr} details our proposed approach, elucidating the methodology for utilizing various LLMs in the context of MGC. This section outlines the models employed, their architectures, and our adaptation strategies. Section \ref{sec:exp} presents and analyzes the experimental results of the LLM-based approaches to MGC. We provide a thorough discussion of our findings, including performance metrics, comparative analyses, and insights derived from our experiments. The final section offers concluding remarks, summarizing the key contributions of our work and proposing directions for future research in this promising intersection of LLMs and MIR.

\section{Related Works}
\label{sec:rel}
The field of MGC has undergone significant evolution over the past two decades, transitioning from traditional expert-driven methods to sophisticated approaches leveraging deep learning and large language models. Initially, MGC relied heavily on the subjective judgments of music industry experts who manually analyzed musical elements. The advent of computing power led to the emergence of automated classification methods. Early approaches to music genre classification relied heavily on engineered features, carefully crafted to capture relevant musical characteristics. These methods typically involved extracting low-level acoustic features such as Mel-frequency cepstral coefficients \cite{Meng2005}, spectral centroid, tempo, timbre, tonality, and rhythmic patterns \cite{Tzanetakis2002,costa2004automatic}. These handcrafted features were then used as inputs to traditional ML algorithms like decision trees \cite{silla2008machine}, k-nearest neighbors \cite{silla2008machine}, multilayer perceptron neural networks \cite{Poitevin2005,silla2008machine}, and support vector machines \cite{Xu2003,silla2008machine}.


The introduction of deep learning marked a significant turning point in MGC research. Deep neural networks, applied to spectrograms and raw waveforms, enabled the automatic extraction of complex features without manual engineering. Convolutional neural networks (CNNs) emerged as powerful tools for automatic feature extraction from audio signals \cite{Dieleman2014music}. 2D CNNs treat spectrograms as images, leveraging techniques from computer vision to capture both temporal and spectral characteristics of music \cite{Dieleman2014music,Han2017music,Choi2017}. Also, 1D CNNs were applied directly to raw audio waveforms or time-frequency representations \cite{Dieleman2014music,Allamy2021}. Recognizing the inherently sequential nature of music, researchers began exploring recurrent neural networks, particularly long short-term memory networks, to capture long-term dependencies in musical sequences \cite{Choi2017}. 

More recently, transformer-based architectures, originally developed for natural language processing tasks, have demonstrated remarkable performance in capturing both local and global contexts in audio. The audio spectrogram transformer (AST) has shown promise for various audio classification tasks \cite{Gong2021}. AST adapts the vision transformer architecture to work with audio spectrograms, allowing it to capture both local and global patterns in audio data. The model treats audio spectrograms as images and applies self-attention mechanisms across the time-frequency representation of audio. Transformers have demonstrated strong performance on tasks like audio event detection and speech recognition, suggesting it could be effective for MGC as well \cite{Harryanto2022}.

Some recent papers discuss the advancements in applying LLMs and foundation models to music understanding and generation \cite{Yuan2023,Li2024,Li2024MERT,Vasilakis2024}. Ma et al.~\cite{Ma2024} provide a comprehensive overview of foundation models applied to music, covering various aspects like model architectures, pre-training methods, and applications. Li et al.~\cite{Li2024} specifically examine foundation models for music comprehension tasks. Li et al.~\cite{Li2024MERT} introduce a new model for acoustic music understanding using self-supervised learning. Vasilakis et al.~\cite{Vasilakis2024} assess the performance of existing language models on music-related tasks. Yuan et al.~\cite{Yuan2023} present a benchmark for evaluating music audio representations across various tasks. Regarding MGC specifically, while these papers do not focus solely on this task, they likely touch on it as part of broader music understanding capabilities.


\section{Proposed Approach}
\label{sec:appr}
This section presents our approach to MCG, which handles variable-length music tracks using audio LLMs in a zero-shot setting. Our approach leverages the power of pre-trained audio LLMs to extract robust and semantically rich features from music tracks, regardless of their duration, and employs a classification head trained on these features for music genre prediction.

\begin{figure*}[htpb!]
  \centering
  \includegraphics[width=0.85\textwidth]{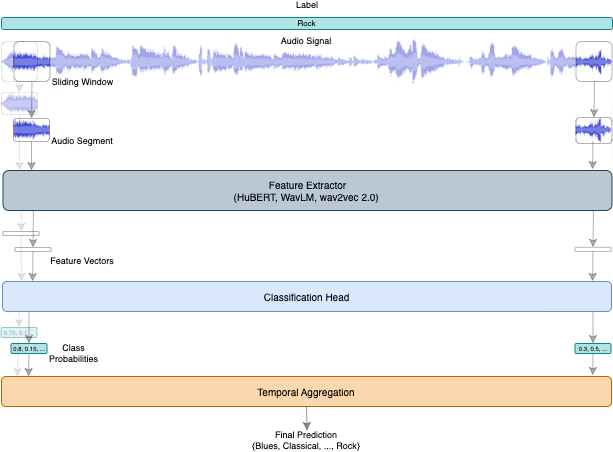}
  \caption{An overview of the proposed approach. The input audio signal is split into 20 ms segments with 25 ms stride. Adapted from~\cite{Koerich2020}.}
  \label{fig:framing}
\end{figure*}

Fig.~\ref{fig:framing} shows an overview of the proposed approach for MGC using audio LLMs as feature extractors in a zero-shot setting. The proposed approach handles music tracks of variable lengths. The audio LLMs have a receptive field of 25 ms, and the shift between consecutive frames is 20 ms. The audio signal has a single label shared with all audio segments during the training of the classification head. Once the classification head is trained, it can be used to predict the music genre of a new music piece. The feature extractor produces a feature vector for each 25 ms audio segment. The prediction is carried out on each audio segment, followed by an aggregating to come up with a final decision on the music genre. This strategy also allows for the prediction of the music genres of audio streams.

Our approach utilizes zero-shot learning for audio LLMs (HuBERT, wav2vec 2.0, and WavLM). This means these models trained on several hours of speech data were not fine-tuned on music data, preserving their generalization capabilities. We only train classification heads consisting of a two-layer perceptron with sigmoid activation at the output layer for multi-class classification.


For the audio LLMs, we have selected WavLM (Base, Base+, and Large), HuBERT Large, and wav2vec 2.0 (Base 960h, Large 960h, and Large+Self-Attention). We extract feature vectors from both the final and intermediate layers of each architecture \cite{Pasad2023}, allowing for a comprehensive analysis of the most suitable representations for MGC. These models require input audio sampled at 16 kHz, and they segment it using a 25 ms window with a 48 ms overlap, resulting in a feature vector every 20 ms. All Large models generate 1024-dimensional vectors. To further refine our analysis, we evaluated all layers of the best-performing LLMs to determine the most effective layers for MGC. We evaluated the 6th, 12th, 18th, and 24th layers for large models and the 3rd, 6th, 9th, and 12th layers for base models.

The extracted feature vectors are fed into a multilayer perceptron (MLP) with two hidden layers of 128 and 64 units, respectively, with ReLU activation function and dropout of 40\% after each layer and a softmax output layer for classification. We trained the model using a batch size of 1500 and the Adam optimizer, measuring both loss and accuracy. Training lasts for up to 200 epochs with early stopping to prevent overfitting.

The model predicts a genre for each audio sample at every 20 ms segment, and a temporal aggregation function such as majority voting, sum rule, weighted voting, and product rule is used to reach a final genre prediction for the entire audio, as illustrated in Fig.~\ref{fig:framing}.


\subsection{Large Language Models (LLMs)}

\subsubsection{HuBERT}
HuBERT is a self-supervised speech representation learning model developed by Facebook AI Research \cite{Hsu2021}. It applies BERT-like masked prediction to continuous inputs, addressing challenges specific to speech: multiple sound units per utterance, no pre-defined lexicon during pre-training, and variable-length units without explicit segmentation. The model architecture consists of a convolutional feature encoder, a transformer encoder, and additional layers for clustering and prediction. HuBERT processes raw audio sampled at 16 kHz, using a 25 ms window with a 20 ms stride to generate feature vectors. HuBERT's training process involves two main steps: (i) offline clustering to create pseudo-labels for masked prediction and (ii) iterative refinement of these clusters using learned representations. HuBERT has achieved state-of-the-art performance on various speech recognition benchmarks, showcasing its effectiveness in learning robust speech representations \cite{Yang2021}.

There are three versions of HuBERT: (i) \textit{HuBERT Base} has 95 million parameters, 12 transformer layers, 8 attention heads, and 768 embedding dimensions, and it was pre-trained on 960 hours of LibriSpeech data; (ii) \textit{HuBERT Large} has 317 million parameters, 24 transformer layers, 16 attention heads, 1024 embedding dimension, and it was pre-trained on 60,000 hours of Libri-Light data; (iii) \textit{HuBERT X-Large} has 964 million parameters, 48 transformer layers, 16 attention heads, 1280 embedding dimension, and it was pre-trained on 60,000 hours of Libri-Light data.

\subsubsection{wav2vec 2.0}
wav2vec 2.0 is a self-supervised learning model for speech recognition developed by Facebook AI Research \cite{Baevski2020}. It processes raw audio waveforms sampled at 16 kHz, extracting 20 ms-long audio representations. These representations are then quantized into a finite set of discrete vectors, creating a codebook of speech units. The model's architecture combines elements from BERT and contrastive learning: (i) it applies masking techniques similar to BERT to capture contextual information; (ii) it uses a contrastive task to learn speech representations, distinguishing between true quantized latents and distractors. This approach allows wav2vec 2.0 to leverage large amounts of unlabeled speech data, significantly reducing the need for manually transcribed audio. Wav2vec 2.0 has demonstrated remarkable performance. When trained on the full Librispeech dataset, it achieves a word error rate (WER) of 1.8\% on clean test sets and 3.3\% on other sets. Pre-training on 53,000 hours of unlabeled data and using only 10 minutes of labeled data, the wav2vec 2.0 model achieved a WER of 4.8\% on clean test sets and 8.2\% on others \cite{Baevski2020}. 

There are different variants of wav2vec 2.0 Base: (i) \textit{wav2vec 2.0 Large} contains 317 million parameters with 24 transformer layers, 16 attention heads, a model dimension of 1024, and is pre-trained on 960 hours of unlabeled LibriSpeech automatic speech recognition (ASR) data; (ii) \textit{wav2vec 2.0 Large-LV60} was pre-trained on 60,000 hours of unlabeled LibriVox data; (iii) \textit{wav2vec 2.0 Large-960h} was pre-trained and fine-tuned on 960 hours of unlabeled LibriSpeech data; (iv) \textit{wav2vec 2.0 Large-960h-lv60-self} was pre-trained on 960 hours of unlabeled LibriSpeech and Libri-Light data, and 60,000 hours of unlabeled LibriVox data, using self-supervision for learning. These variants offer different trade-offs between model size, training data, and performance, allowing researchers to choose the most suitable version for their specific tasks and computational resources.

\subsubsection{WavLM}
WavLM \cite{Chen2022} is a self-supervised learning model developed by Microsoft for comprehensive speech processing tasks, including ASR, speaker recognition, speech separation, and speaker diarization. Building upon the HuBERT framework, WavLM predicts masked portions of input speech while incorporating noise reduction techniques. Key features of WavLM include a transformer structure enhanced with gated relative position bias to effectively capture the sequential order of input data. WavLM accepts raw waveforms sampled at 16 kHz, extracting audio representations using a 20 ms window, and employs temporal masking and contrastive learning methods similar to BERT \cite{Chen2022}. WavLM was trained on a 94,000-hour dataset comprising Libri-Light, GigaSpeech, and VoxPopuli. WavLM has demonstrated state-of-the-art performance on the SUPERB benchmark, surpassing HuBERT and wav2vec 2.0 on several subtasks \cite{Yang2021}.


There are three versions of the WavLM model: (i) \textit{WavLM Base} has 94.7 million parameters with 12 transformer layers, 8 attention heads, a model dimension of 768, and is pre-trained on 960 hours of unlabeled LibriSpeech data; (ii) \textit{WavLM Base+} has 94.7 million parameters with 12 transformer layers, 8 attention heads, a model dimension of 768, and is pre-trained on 10,000 hours of GigaSpeech, 60,000 hours of Libri-Light, and 24,000 hours of VoxPopuli unlabeled data; (iii) \textit{WavLM Large} has 316.6 million parameters with 24 transformer layers, 12 attention heads, a model dimension of 1024, and is pre-trained on 10,000 hours of GigaSpeech, 60,000 hours of Libri-Light, and 24,000 hours of VoxPopuli unlabeled data.

\section{Experimental Results}
\label{sec:exp}
This section presents a comprehensive evaluation of our proposed approach for MGC. We begin by describing the dataset used in our experiments and detailing our experimental protocol. Following this, we report the results of extensive experiments conducted with HuBERT, WavLM, and wav2vec 2.0, representing the current state-of-the-art in speech and audio processing. To provide a thorough assessment of our method's performance, we compare these results with those obtained from transformer-based models and CNN architectures, which have shown significant promise in various audio classification tasks.

\subsection{Dataset}
The GTzan dataset is the most widely used public resource for MGC research, and it encompasses various recording conditions from various sources, including radio broadcasts, personal CDs, and microphone recordings \cite{Tzanetakis2002}. The dataset comprises 1,000 30-second music excerpts distributed across ten genres: classical, rock, blues, reggae, metal, disco, country, jazz, hip-hop, and pop \cite{Tzanetakis2002}. Each genre contains 100 30-second audio clips, totaling approximately five hours of audio data in WAV format, originally sampled at 22,050 Hz with 16-bit resolution. Despite its popularity, GTzan has known integrity issues, including duplicates, incorrect labels, and distortions. These problems have been extensively documented and should be considered when interpreting results based on this dataset.

For our experiment, we resampled the GTzan dataset to 16 kHz to ensure compatibility with the audio LLMs, which are optimized for this sampling rate. It is important to note that resampling can affect the accuracy of automatic MGC, leading to data loss, particularly in high-frequency content. This loss can potentially impact the algorithm's ability to identify certain genres, especially those with prominent high-frequency components like electronic or metal music.

To implement a 3-fold cross-validation protocol, we shuffled and divided the 1,000 recordings into three sets of approximately 334, 333, and 333 samples. This approach allows us to use each set as a distinct dataset for training, validation, and testing, enhancing the robustness of our evaluation. We primarily used PyTorch, Pandas, NumPy, and the Transformers library for our implementation\footnote{The source code available at: \url{https://github.com/megamine25/Music-genre-classification}}.


\subsection{Results for LLMs}
Table~\ref{tab:llms} presents the average accuracy and standard deviation achieved by different versions of WavLM, wav2vec 2.0, and HuBERT models on the 3-fold cross-validation of the GTzan dataset. The WavLM Large model consistently outperformed other audio LLM models, prompting a more in-depth evaluation of its intermediate layer representations.
The WavLM Large model demonstrated superior performance, achieving 75.5\% accuracy on segments at the 5th layer and 84.6\% accuracy after aggregation at the 11th layer. This exceptional performance can be attributed to the model's ability to handle diverse language and audio processing tasks, capturing complex audio features and learning rich representations at various depth levels. The diversity of intermediate layers allows the model to grasp both local and global information, enhancing its ability to differentiate music genres accurately and robustly.

When evaluating the intermediate layers of other audio LLM models, a common trend emerged: higher accuracy in the earlier layers, followed by decreasing accuracy in deeper layers. This pattern is likely due to the models' primary design for speech recognition tasks. In earlier layers, these models capture low-level features, such as fundamental frequencies and harmonic structures, which are relevant for MGC. However, deeper layers focus more on high-level features specific to speech recognition, such as linguistic structure and phonetic nuances, which are less relevant for MGC, resulting in decreased accuracy.


\begin{table}[h!]
    \centering
     \caption{Average accuracy and standard deviation achieved by different versions of LLM models on 3-fold of the GTzan dataset. The best results for each model on segments and the whole music track are underlined.}
     \setlength{\tabcolsep}{5pt}
     \renewcommand{\arraystretch}{1.2}
    \begin{tabular}{|l|l|c|r|r|}
        \hline
        &  &  & \multicolumn{2}{c|}{\textbf{Mean$\pm$StD (\%)}} \\
        \cline{4-5}
        \textbf{Model} & \textbf{Version} & \textbf{Layer} & \multicolumn{1}{c|}{\textbf{Segments}} & \multicolumn{1}{c|}{\textbf{Aggregation}} \\
        \hline
        \multirow{4}{*}{HuBERT}  & \multirow{4}{*}{Large} & 6 & \underline{73.216$\pm$1.702} & \underline{81.400$\pm$0.901} \\
        & & 12 & 68.077$\pm$1.160 & 80.902$\pm$3.018 \\
        & & 18 & 59.253$\pm$1.424 & 71.800$\pm$0.812 \\
        & & 24 & 62.104$\pm$2.165 & 73.197$\pm$3.489 \\
        \hline
 \multirow{12}{*}{wav2vec 2.0} & \multirow{4}{*}{Large+ST} & 6 & \underline{72.741$\pm$1.193} & \underline{81.199$\pm$1.672} \\
        &  & 12 & 63.749$\pm$1.844 & 79.599$\pm$1.578 \\
        &  & 18 & 62.759$\pm$1.769 & 78.000$\pm$1.550 \\
        &  & 24 & 53.889$\pm$2.068 & 68.099$\pm$3.222 \\
          \cline{2-5}
        & \multirow{4}{*}{Large 960h}  & 6 & 66.713$\pm$2.150 & 78.199$\pm$2.008 \\
        &   & 12 & 59.002$\pm$1.220 & 73.600$\pm$1.385 \\
        &   & 18 & 54.725$\pm$0.881 & 71.301$\pm$1.452 \\
        &   & 24 & 39.772$\pm$0.823 & 46.501$\pm$0.600 \\
           \cline{2-5}
        & \multirow{4}{*}{Base 960h}  & 3 & 67.311$\pm$1.404 & 76.799$\pm$1.784 \\
        &  & 6 & 60.094$\pm$2.258 & 71.499$\pm$4.432 \\
        &   & 9 & 53.222$\pm$0.600 & 68.602$\pm$1.873 \\
        &   & 12 & 42.162$\pm$1.549 & 55.100$\pm$1.952 \\
        \hline
        \multirow{32}{*}{WavLM} & \multirow{24}{*}{Large} & 1 & 51.067$\pm$0.997 & 73.097$\pm$2.591 \\
        &  & 2 & 68.666$\pm$2.040 & 74.897$\pm$2.588 \\
        &  & 3 & 73.312$\pm$1.554 & 79.999$\pm$2.595 \\
        &  & 4 & 75.219$\pm$1.380 & 81.900$\pm$2.407 \\
        &  & 5 & \underline{75.506$\pm$1.716} & 83.099$\pm$2.284 \\
        &  & 6 & 74.384$\pm$1.715 & 83.598$\pm$2.960 \\
        &  & 7 & 74.038$\pm$2.015 & 82.697$\pm$3.770 \\
        &  & 8 & 73.154$\pm$1.697 & 83.398$\pm$2.533 \\
        &  & 9 & 72.438$\pm$0.886 & 82.899$\pm$2.464 \\
        &  & 10 & 72.723$\pm$1.321 & 83.700$\pm$1.258 \\
        &  & 11 & 72.350$\pm$1.109 & \underline{84.600$\pm$2.135} \\
        &  & 12 & 71.909$\pm$1.552 & 84.101$\pm$2.091 \\
        &  & 13 & 70.788$\pm$1.327 & 82.900$\pm$2.102 \\
        &  & 14 & 68.424$\pm$1.746 & 82.300$\pm$1.667 \\
        &  & 15 & 66.275$\pm$1.771 & 81.399$\pm$1.327 \\
        &  & 16 & 64.847$\pm$1.613 & 80.698$\pm$3.321 \\
        &  & 17 & 64.077$\pm$1.416 & 79.900$\pm$1.502 \\
        &  & 18 & 64.082$\pm$1.601 & 80.100$\pm$2.432 \\
        &  & 19 & 64.292$\pm$1.366 & 80.100$\pm$1.357 \\
        &  & 20 & 64.311$\pm$1.123 & 79.600$\pm$1.202 \\
        &  & 21 & 64.090$\pm$0.720 & 79.300$\pm$1.202 \\
        &  & 22 & 66.120$\pm$0.757 & 80.201$\pm$2.848 \\
        &  & 23 & 66.120$\pm$0.757 & 80.201$\pm$2.848 \\
        &  & 24 & 65.044$\pm$1.300 & 79.100$\pm$2.720 \\
        \cline{2-5}
        & \multirow{4}{*}{Base+}  & 3 & 68.513$\pm$1.562 & 79.400$\pm$1.352 \\
        &  & 6 & 67.042$\pm$2.100 & 79.100$\pm$2.253 \\
        &   & 9 & 61.617$\pm$1.567 & 74.900$\pm$2.553 \\
        &   & 12 & 63.213$\pm$1.439 & 72.601$\pm$2.403 \\
        \cline{2-5}
        & \multirow{4}{*}{Base}  & 3 & 68.088$\pm$1.794 & 77.700$\pm$3.156 \\
        &   & 6 & 60.588$\pm$2.123 & 76.399$\pm$1.677 \\
        &   & 9 & 52.553$\pm$1.498 & 71.400$\pm$0.719 \\
        &   & 12 & 56.749$\pm$1.217 & 67.801$\pm$1.359 \\
        \hline
    \end{tabular}
    \label{tab:llms}
\end{table}

Table~\ref{tab:comp} presents a comprehensive comparison of the best-performing versions of each LLM from Table~\ref{tab:llms} with various deep learning architectures, including 1D and 2D CNNs and an audio transformer. We have chosen the AST model, which is the first purely attention-based model for audio classification supporting variable length input that can be applied to different tasks \cite{Gong2021}. The AST was also used in a zero-shot approach as LLMs, and only the representation produced at its last layer (12th) was evaluated. This comparison is also based on the average accuracy and standard deviation achieved on the 3-fold cross-validation of the GTzan dataset. Table~\ref{tab:comp} provides insights into the relative performance of different model types in the context of MGC, allowing for a nuanced understanding of their strengths and limitations.

\begin{table}[h!]
    \centering
     \caption{Average accuracy and standard deviation achieved by different deep architectures on 3-fold of the GTzan dataset.}
    \renewcommand{\arraystretch}{1.3}
    \begin{tabular}{|l|r|r|}
        \hline
        &    \multicolumn{2}{c|}{\textbf{Mean$\pm$StD (\%)}} \\
        \cline{2-3}
        \multicolumn{1}{|c|}{\textbf{Model}}  & \multicolumn{1}{c|}{\textbf{Segments}} & \multicolumn{1}{c|}{\textbf{Aggregation}} \\
        \hline
        {WavLM} Large { (Layer 11)} & 72.350$\pm$1.109 & 84.600$\pm$2.135 \\
        {WavLM} Base+ {\hspace{1pt}(Layer 3)} & 68.513$\pm$1.562 & 79.400$\pm$1.352 \\
        {WavLM} Base  {\hspace{5pt}(Layer 3)}&  68.088$\pm$1.794 & 77.700$\pm$3.156 \\
        \hline
        {wav2vec 2.0} {Large+ST} {\hspace{5pt}(Layer 6)} & 72.741$\pm$1.193 & 81.199$\pm$1.672 \\
        {wav2vec 2.0} {Large 960h} {(Layer 6)}  & 66.713$\pm$2.150 & 78.199$\pm$2.008 \\
        {wav2vec 2.0} {Base 960h} { (Layer 3)} & 67.311$\pm$1.404 & 76.799$\pm$1.784 \\
        \hline
        {HuBERT} {Large} {(Layer 6)} & 73.216$\pm$1.702 & 81.400$\pm$0.901 \\
        \hline
        {AST} {(Layer 12)}  & 79.752$\pm$1.439 & 85.500$\pm$1.840 \\
        \hline
		1D CRNN \cite{Choi2017} & 71.230$\pm$1.030 & 79.320$\pm$1.830 \\
        1D CNN \cite{Pons2018} & 66.310$\pm$1.350 &  74.320$\pm$1.640 \\
        1D CNN \cite{Dieleman2014} & 45.200$\pm$2.220 & 53.620$\pm$2.400 \\
		1D CNN \cite{abdoli2019} & 15.000$\pm$3.860 & 15.590$\pm$4.170 \\
		1D CNN \cite{Koerich2020} & 64.900$\pm$1.730 & 73.020$\pm$1.100 \\
		1D ResNet CNN \cite{Allamy2021} & 74.620$\pm$1.910 & 80.930$\pm$2.350 \\
        \hline
        2D STFT CNN \cite{Koerich2020} & 75.290$\pm$2.500 & 81.370$\pm$2.200 \\
        \hline
    \end{tabular}
    \label{tab:comp}
\end{table}

\subsection{Discussion}
Our study reveals several key insights into the performance of various models for MGC. The AST model outperformed all other models in our study, achieving 79.75\% accuracy on audio segments and 85.50\% after aggregation. This result underscores the effectiveness of attention mechanisms in capturing complex temporal and frequency features from audio signals. On the other hand, LLMs, particularly WavLM Large, demonstrated strong capabilities in audio feature extraction. The exceptional performance of WavLM Large's 5th and 11th internal layers highlights the potential of these models for music-related tasks. We observed that, in general, the initial layers of LLM models generally exhibited superior performance in capturing audio and musical features, resulting in higher classification accuracy. This suggests that these layers may be more adept at identifying low-level audio characteristics relevant to genre classification. 

Transformer-based models, AST, and LLMs consistently outperformed traditional approaches like 1D and 2D CNNs. This indicates a significant advancement in the field of MIR. While these findings are promising, it's important to acknowledge certain limitations of our study. The GTzan dataset, despite its widespread use, has known integrity issues and limited genre diversity. Additionally, variability in audio recordings and potential biases in the dataset may affect the models' generalization capabilities.

Nevertheless, our results have significant implications for the MIR domain. The superior performance of transformer-based models and LLMs opens new avenues for improving MGC systems and potentially other music-related tasks. Future research could explore fine-tuning these models on larger, more diverse datasets or investigating their effectiveness in related tasks such as music recommendation or mood classification.


The GTzan dataset is a popular benchmark for MGC research. Our proposed approach using WavLM, wav2vec 2.0, and HuBERT achieved accuracies of 84.6\%, 81.2\%, and 81.4\%, respectively, while the AST achieved 85.5\%. While these results are promising, it is important to note that direct comparisons with other studies can be challenging due to variations in experimental setups and test splits. Reported accuracies on the GTzan dataset vary widely, typically ranging from 70\% to over 90\%. However, the highest reported accuracies should be interpreted cautiously, as they might be influenced by overfitting. Our experimental protocol mitigated this risk by employing random splits and 3-fold cross-validation. 

While the proposed approach based on LLMs in their zero-shot capacity demonstrated promising results, their performance could be further improved by fine-tuning them on a music dataset. This would allow the models to learn the specific representations of music data, potentially leading to more accurate and informative feature representations.

\section{Conclusions}
Our comprehensive analysis of various deep learning architectures for MGC has yielded several significant insights: (i) The WavLM Large model demonstrated superior performance among LLM architectures, while the transformer-based AST model achieved the highest overall accuracy. This highlights the potential of transformer architectures in MIR tasks; (ii) The initial layers of audio LLM models proved particularly effective in capturing musical features, suggesting that these layers may be more adept at identifying low-level audio characteristics relevant to genre classification; (iii) The zero-shot approach for AST and LLMs, contrasted with fully-trained CNNs, provides valuable insights into the generalization capabilities of these models in the music domain. However, we acknowledge certain limitations: (i) The dataset's limited genre diversity and potential integrity issues may affect model generalization; (ii) The zero-shot approach, while demonstrating the models' adaptability, may not fully exploit their potential in music-specific tasks.

Future research directions could address these limitations and further advance the field, such as implementing cross-attention mechanisms between different models, which could enhance performance by allowing better inter and intra-model interaction \cite{Praveen2022}. Besides, utilizing larger, more diverse datasets such as the Free Music Archive, ISMIR2004, or the Latin music database \cite{silla2008latin} could improve model robustness and generalization. Also, exploring the impact of fine-tuning audio LLMs and transformers on music data could potentially boost their performance in genre classification tasks.

In conclusion, our study has made important contributions to MGC while opening new avenues for future research. By addressing the identified limitations and exploring proposed improvements, future work can further enhance the accuracy and efficiency of classification models in the MIR domain.

\section*{Acknowledgment}
This work was funded by the Natural Sciences and Engineering Research Council of Canada (NSERC) under Grant RGPIN-2022-05074.

\newpage
\clearpage
\balance

\bibliographystyle{IEEEtran}
\bibliography{IEEEabrv,MoodNet,llmaudio}

\begin{thebibliography}{10}
\providecommand{\url}[1]{#1}
\csname url@samestyle\endcsname
\providecommand{\newblock}{\relax}
\providecommand{\bibinfo}[2]{#2}
\providecommand{\BIBentrySTDinterwordspacing}{\spaceskip=0pt\relax}
\providecommand{\BIBentryALTinterwordstretchfactor}{4}
\providecommand{\BIBentryALTinterwordspacing}{\spaceskip=\fontdimen2\font plus
\BIBentryALTinterwordstretchfactor\fontdimen3\font minus
  \fontdimen4\font\relax}
\providecommand{\BIBforeignlanguage}[2]{{%
\expandafter\ifx\csname l@#1\endcsname\relax
\typeout{** WARNING: IEEEtran.bst: No hyphenation pattern has been}%
\typeout{** loaded for the language `#1'. Using the pattern for}%
\typeout{** the default language instead.}%
\else
\language=\csname l@#1\endcsname
\fi
#2}}
\providecommand{\BIBdecl}{\relax}
\BIBdecl

\bibitem{Lidy2010}
T.~Lidy, C.~N.~S. Jr., O.~Cornelis, F.~Gouyon, A.~Rauber, C.~A.~A. Kaestner,
  and A.~L. Koerich, ``On the suitability of state-of-the-art music information
  retrieval methods for analyzing, categorizing and accessing non-western and
  ethnic music collections,'' \emph{Signal Process.}, vol.~90, no.~4, pp.
  1032--1048, 2010.

\bibitem{REBECCHINI2021}
L.~Rebecchini, ``Music, mental health, and immunity,'' \emph{Brain, Behavior,
  \& Immunity - Health}, vol.~18, p. 100374, 2021.

\bibitem{Prabhakar2023}
S.~K. Prabhakar and S.~Lee, ``Holistic approaches to music genre classification
  using efficient transfer and deep learning techniques,'' \emph{Expert Syst.
  Appl.}, vol. 211, p. 118636, 2023.

\bibitem{Tzanetakis2002}
G.~Tzanetakis and P.~R. Cook, ``Musical genre classification of audio
  signals,'' \emph{{IEEE} Trans Speech Audio Process}, vol.~10, no.~5, pp.
  293--302, 2002.

\bibitem{Schedl2014}
M.~Schedl, E.~G{\'{o}}mez, and J.~Urbano, ``Music information retrieval: Recent
  developments and applications,'' \emph{Found. Trends Inf. Retr.}, vol.~8, no.
  2-3, pp. 127--261, 2014.

\bibitem{Burgoyne2015}
J.~A. Burgoyne, I.~Fujinaga, and J.~S. Downie, \emph{Music Information
  Retrieval}.\hskip 1em plus 0.5em minus 0.4em\relax John Wiley \& Sons, Ltd,
  2015, ch.~15, pp. 213--228.

\bibitem{Allamy2021}
S.~Allamy and A.~Lameiras, ``{1D CNN} architectures for music genre
  classification,'' in \emph{{IEEE} Symposium Series on Computational
  Intelligence, {SSCI} 2021, Orlando, FL, USA, December 5-7, 2021}.\hskip 1em
  plus 0.5em minus 0.4em\relax {IEEE}, 2021, pp. 1--7.

\bibitem{Yang2021}
S.~Yang, P.~Chi, Y.~Chuang, C.~J. Lai, K.~Lakhotia, Y.~Y. Lin, A.~T. Liu,
  J.~Shi, X.~Chang, G.~Lin, T.~Huang, W.~Tseng, K.~Lee, D.~Liu, Z.~Huang,
  S.~Dong, S.~Li, S.~Watanabe, A.~Mohamed, and H.~Lee, ``{SUPERB:} speech
  processing universal performance benchmark,'' in \emph{22nd Annual Conference
  of the International Speech Communication Association, Interspeech 2021,
  Brno, Czechia, August 30 - September 3, 2021}, H.~Hermansky,
  H.~Cernock{\'{y}}, L.~Burget, L.~Lamel, O.~Scharenborg, and
  P.~Motl{\'{\i}}cek, Eds., 2021, pp. 1194--1198.

\bibitem{Hsu2021}
W.~Hsu, B.~Bolte, Y.~H. Tsai, K.~Lakhotia, R.~Salakhutdinov, and A.~Mohamed,
  ``Hubert: Self-supervised speech representation learning by masked prediction
  of hidden units,'' \emph{{IEEE} {ACM} Trans. Audio Speech Lang. Process.},
  vol.~29, pp. 3451--3460, 2021.

\bibitem{Baevski2020}
A.~Baevski, Y.~Zhou, A.~Mohamed, and M.~Auli, ``wav2vec 2.0: {A} framework for
  self-supervised learning of speech representations,'' in \emph{Advances in
  Neural Information Processing Systems 33: Annual Conference on Neural
  Information Processing Systems 2020, NeurIPS 2020, December 6-12, 2020,
  virtual}, H.~Larochelle, M.~Ranzato, R.~Hadsell, M.~Balcan, and H.~Lin, Eds.,
  2020.

\bibitem{Chen2022}
S.~Chen, C.~Wang, Z.~Chen, Y.~Wu, S.~Liu, Z.~Chen, J.~Li, N.~Kanda,
  T.~Yoshioka, X.~Xiao, J.~Wu, L.~Zhou, S.~Ren, Y.~Qian, Y.~Qian, J.~Wu,
  M.~Zeng, X.~Yu, and F.~Wei, ``Wavlm: Large-scale self-supervised pre-training
  for full stack speech processing,'' \emph{{IEEE} J. Sel. Top. Signal
  Process.}, vol.~16, no.~6, pp. 1505--1518, 2022.

\bibitem{Pasad2023}
A.~Pasad, B.~Shi, and K.~Livescu, ``Comparative layer-wise analysis of
  self-supervised speech models,'' in \emph{{IEEE} International Conference on
  Acoustics, Speech and Signal Processing {ICASSP} 2023, Rhodes Island, Greece,
  June 4-10, 2023}, 2023, pp. 1--5.

\bibitem{Meng2005}
A.~Meng, P.~Ahrendt, and J.~Larsen, ``Improving music genre classification by
  short time feature integration,'' in \emph{2005 {IEEE} International
  Conference on Acoustics, Speech, and Signal Processing, {ICASSP} '05,
  Philadelphia, Pennsylvania, USA, March 18-23, 2005}, 2005, pp. 497--500.

\bibitem{costa2004automatic}
C.~H.~L. Costa, J.~D. Valle, and A.~L. Koerich, ``Automatic classification of
  audio data,'' in \emph{{IEEE} Intl Conf Syst, Man, Cybernetics}, 2004, pp.
  562--567.

\bibitem{silla2008machine}
C.~N. Silla, A.~L. Koerich, and C.~A.~A. Kaestner, ``A machine learning
  approach to automatic music genre classification,'' \emph{Journal of the
  Brazilian Computer Society}, vol.~14, no.~3, pp. 7--18, 2008.

\bibitem{Poitevin2005}
C.~Poitevin and A.~Lameiras~Koerich, ``Combination of homogeneous classifiers
  for musical genre classification,'' in \emph{2005 IEEE International
  Conference on Systems, Man and Cybernetics}, 2005, pp. 554--559.

\bibitem{Xu2003}
C.~Xu, N.~Maddage, X.~Shao, F.~Cao, and Q.~Tian, ``Musical genre classification
  using support vector machines,'' in \emph{2003 IEEE International Conference
  on Acoustics, Speech, and Signal Processing, 2003. Proceedings. (ICASSP
  '03).}, vol.~5, 2003, pp. V--429.

\bibitem{Dieleman2014music}
S.~{Dieleman} and B.~{Schrauwen}, ``End-to-end learning for music audio,'' in
  \emph{2014 IEEE International Conference on Acoustics, Speech and Signal
  Processing (ICASSP)}, 2014, pp. 6964--6968.

\bibitem{Han2017music}
Y.~{Han}, J.~{Kim}, and K.~{Lee}, ``Deep convolutional neural networks for
  predominant instrument recognition in polyphonic music,'' \emph{IEEE/ACM
  Transactions on Audio, Speech, and Language Processing}, vol.~25, no.~1, pp.
  208--221, 2017.

\bibitem{Choi2017}
K.~Choi, G.~Fazekas, M.~B. Sandler, and K.~Cho, ``Convolutional recurrent
  neural networks for music classification,'' in \emph{2017 {IEEE}
  International Conference on Acoustics, Speech and Signal Processing, {ICASSP}
  2017, New Orleans, LA, USA, March 5-9, 2017}, 2017, pp. 2392--2396.

\bibitem{Gong2021}
Y.~Gong, Y.-A. Chung, and J.~Glass, ``{AST: Audio Spectrogram Transformer},''
  in \emph{Proc. Interspeech 2021}, 2021, pp. 571--575.

\bibitem{Harryanto2022}
A.~A.~A. Harryanto, K.~Gunawan, R.~Nagano, and R.~Sutoyo, ``Music
  classification model development based on audio recognition using transformer
  model,'' in \emph{3rd International Conference on Artificial Intelligence and
  Data Sciences (AiDAS)}, 2022, pp. 258--263.

\bibitem{Yuan2023}
R.~Yuan, Y.~Ma, Y.~Li, G.~Zhang, X.~Chen, H.~Yin, z.~le, Y.~Liu, J.~Huang,
  Z.~Tian, B.~Deng, N.~Wang, C.~Lin, E.~Benetos, A.~Ragni, N.~Gyenge,
  R.~Dannenberg, W.~Chen, G.~Xia, W.~Xue, S.~Liu, S.~Wang, R.~Liu, Y.~Guo, and
  J.~Fu, ``Marble: Music audio representation benchmark for universal
  evaluation,'' in \emph{Advances in Neural Information Processing Systems},
  A.~Oh, T.~Naumann, A.~Globerson, K.~Saenko, M.~Hardt, and S.~Levine, Eds.,
  vol.~36.\hskip 1em plus 0.5em minus 0.4em\relax Curran Associates, Inc.,
  2023, pp. 39\,626--39\,647.

\bibitem{Li2024}
W.~Li, Y.~Cai, Z.~Wu, W.~Zhang, Y.~Chen, R.~Qi, M.~Dong, P.~Chen, X.~Dong,
  F.~Shi, L.~Guo, J.~Han, B.~Ge, T.~Liu, L.~Gan, and T.~Zhang, ``A survey of
  foundation models for music understanding,'' 2024.

\bibitem{Li2024MERT}
Y.~Li, R.~Yuan, G.~Zhang, Y.~Ma, X.~Chen, H.~Yin, C.~Xiao, C.~Lin, A.~Ragni,
  E.~Benetos, N.~Gyenge, R.~B. Dannenberg, R.~Liu, W.~Chen, G.~Xia, Y.~Shi,
  W.~Huang, Z.~Wang, Y.~Guo, and J.~Fu, ``{MERT:} acoustic music understanding
  model with large-scale self-supervised training,'' in \emph{The Twelfth
  International Conference on Learning Representations, {ICLR} 2024, Vienna,
  Austria, May 7-11, 2024}, 2024.

\bibitem{Vasilakis2024}
Y.~Vasilakis, R.~Bittner, and J.~Pauwels, ``Evaluation of pretrained language
  models on music understanding,'' 2024.

\bibitem{Ma2024}
Y.~Ma, A.~Øland, A.~Ragni, B.~M.~D. Sette, C.~Saitis, C.~Donahue, C.~Lin,
  C.~Plachouras, E.~Benetos, E.~Shatri, F.~Morreale, G.~Zhang, G.~Fazekas,
  G.~Xia, H.~Zhang, I.~Manco, J.~Huang, J.~Guinot, L.~Lin, L.~Marinelli,
  M.~W.~Y. Lam, M.~Sharma, Q.~Kong, R.~B. Dannenberg, R.~Yuan, S.~Wu, S.-L. Wu,
  S.~Dai, S.~Lei, S.~Kang, S.~Dixon, W.~Chen, W.~Huang, X.~Du, X.~Qu, X.~Tan,
  Y.~Li, Z.~Tian, Z.~Wu, Z.~Wu, Z.~Ma, and Z.~Wang, ``Foundation models for
  music: A survey,'' 2024.

\bibitem{Koerich2020}
K.~M. {Koerich}, M.~{Esmaeilpour}, S.~{Abdoli}, A.~S. {Britto Jr.}, and A.~L.
  {Koerich}, ``Cross-representation transferability of adversarial attacks:
  From spectrograms to audio waveforms,'' in \emph{Intl Joint Conf on Neural
  Networks}, 2020, pp. 1--7.

\bibitem{Pons2018}
J.~Pons, O.~Nieto, M.~Prockup, E.~Schmidt, A.~Ehmann, and X.~Serra,
  ``End-to-end learning for music audio tagging at scale,'' in \emph{Intl
  Society for Music Inf Retrieval Conf}, 2018, pp. 1--8.

\bibitem{Dieleman2014}
S.~Dieleman and B.~Schrauwen, ``End-to-end learning for music audio,'' in
  \emph{IEEE Intl Conf on Acoustics, Speech and Signal Process}, 2014, pp.
  6964--6968.

\bibitem{abdoli2019}
S.~Abdoli, P.~Cardinal, and A.~L. Koerich, ``End-to-end environmental sound
  classification using a 1d convolutional neural network,'' \emph{Expert
  Systems with Applications}, vol. 136, pp. 252--263, 2019.

\bibitem{Praveen2022}
R.~G. Praveen, W.~C. de~Melo, N.~Ullah, H.~Aslam, O.~Zeeshan, T.~Denorme,
  M.~Pedersoli, A.~L. Koerich, S.~Bacon, P.~Cardinal, and E.~Granger, ``A joint
  cross-attention model for audio-visual fusion in dimensional emotion
  recognition,'' in \emph{{IEEE/CVF} Conference on Computer Vision and Pattern
  Recognition Workshops, {CVPR} Workshops 2022, New Orleans, LA, USA, June
  19-20, 2022}, 2022, pp. 2485--2494.

\bibitem{silla2008latin}
C.~N. Silla~Jr., A.~L. Koerich, and C.~A.~A. Kaestner, ``The latin music
  database,'' in \emph{Intl Society for Music Inf Retrieval Conf}.\hskip 1em
  plus 0.5em minus 0.4em\relax Philadelphia, USA, 2008, pp. 451--456.

\end{thebibliography}


\begin{thebibliography}{00}
\bibitem{b1} G. Eason, B. Noble, and I. N. Sneddon, ``On certain integrals of Lipschitz-Hankel type involving products of Bessel functions,'' Phil. Trans. Roy. Soc. London, vol. A247, pp. 529--551, April 1955.
\bibitem{b2} J. Clerk Maxwell, A Treatise on Electricity and Magnetism, 3rd ed., vol. 2. Oxford: Clarendon, 1892, pp.68--73.
\bibitem{b3} I. S. Jacobs and C. P. Bean, ``Fine particles, thin films and exchange anisotropy,'' in Magnetism, vol. III, G. T. Rado and H. Suhl, Eds. New York: Academic, 1963, pp. 271--350.
\bibitem{b4} K. Elissa, ``Title of paper if known,'' unpublished.
\bibitem{b5} R. Nicole, ``Title of paper with only first word capitalized,'' J. Name Stand. Abbrev., in press.
\bibitem{b6} Y. Yorozu, M. Hirano, K. Oka, and Y. Tagawa, ``Electron spectroscopy studies on magneto-optical media and plastic substrate interface,'' IEEE Transl. J. Magn. Japan, vol. 2, pp. 740--741, August 1987 [Digests 9th Annual Conf. Magnetics Japan, p. 301, 1982].
\bibitem{b7} M. Young, The Technical Writer's Handbook. Mill Valley, CA: University Science, 1989.
\bibitem{b8} D. P. Kingma and M. Welling, ``Auto-encoding variational Bayes,'' 2013, arXiv:1312.6114. [Online]. Available: https://arxiv.org/abs/1312.6114
\bibitem{b9} S. Liu, ``Wi-Fi Energy Detection Testbed (12MTC),'' 2023, gitHub repository. [Online]. Available: https://github.com/liustone99/Wi-Fi-Energy-Detection-Testbed-12MTC
\bibitem{b10} ``Treatment episode data set: discharges (TEDS-D): concatenated, 2006 to 2009.'' U.S. Department of Health and Human Services, Substance Abuse and Mental Health Services Administration, Office of Applied Studies, August, 2013, DOI:10.3886/ICPSR30122.v2
\bibitem{b11} K. Eves and J. Valasek, ``Adaptive control for singularly perturbed systems examples,'' Code Ocean, Aug. 2023. [Online]. Available: https://codeocean.com/capsule/4989235/tree
\end{thebibliography}


\end{document}